\documentclass[journal,transmag]{IEEEtran}

\usepackage{graphicx}
\usepackage{amsmath}

\ifCLASSINFOpdf
\else
\fi
\begin{document}
\title{A Survey on Data Processing Methods and Cloud Computation}

\author{\IEEEauthorblockN{Katherine Hughes\IEEEauthorrefmark{1}
}

}

\IEEEtitleabstractindextext{%
\begin{abstract}

As new technologies move to the fore, our understanding of the world may seem to have shrunk in comparison, for despite new developments in research, much of it is reduced or rather, abstracted for marketability. Thus, the purpose of this work is to provide a brief study of current techniques used in data analysis that will yield some clarity and encourage new research. Several applications of which are discussed in relation to the high data volume fields e.g. health care and economics. Particular attention will be given to recent developments in health care.

\end{abstract}

}

\maketitle

\IEEEdisplaynontitleabstractindextext

\IEEEpeerreviewmaketitle

\section{Introduction}

\IEEEPARstart Repositories of information have existed for thousands of years. As such, the shift from stone tablets to electric fields should seem to make little to no difference so long as the intent to preserve, and record knowledge remains the persists unadulterated. However, much has changed with the advent of data mining and big data computation. For rather than obtaining knowledge through a gradual absorption of meaning there are now methods of research that are a form of information distillation, quite apart from past works. Granted, the results of these inquiries must still be shifted from data to information, much of the process is obfuscated before its introduction to the casual observer. And so we arrive at computation as an abstraction of knowledge. Whether it is the optimization of search engines, exploration of land, sea, and air, or the study of disease data information has become a commodity of scale.  More specifically, in the field of medicine, current data analysis methods have yielded a significant opportunity to improve life for many people with the personalization of medicine. However, incorporating personalization to more than a select group remains problematic particularly for the "...generation and management (storage, and computational resources) of omics data" which "remain expensive despite technological progress" [25]. Additionally, efforts to collect data prior in clinical trials is problematic in that data becomes skewed or corrupt for many reasons such as patient dropout. In other fields, algorithms in machine learning applications of pattern recognition software may result in data that is either partially incomplete or corrupted by noise [26].\\

\section{Cloud Computation and Storage}\label{SectionII}

 Working towards storing and computing data in a cost efficient manner, there are many cloud distributed resources freely available and on a pay-per-use basis. Cloud computation is a broad term that has come to include both public and private as well as hybridization of data cloud resources. Private clouds differ from public systems in that they are tend to be riskier in terms of investment return, whereas resource sharing allows public clouds to have a higher return of capital. The trade off is that public clouds are less secure than their private counterparts. Thus, depending on their needs, many companies may elect to invest in some form of hybrid system i.e. a mixture of public and private cloud. For many companies, their primary concern is reducing costs of companies who attempt to process large volumes of data. Authors in [3] offer "distributed applications" as an alternative to customized processing frameworks (high throughput and many-task computing). Virtual machines are then used to overcome the discrepancy between static cluster environments and the allocation of dynamic resources. As such, a system that allocates resources dynamically for task scheduling and execution, is provided.

 Similarly, the authors of [4] attempt to improve resource allocation in clouds however they focus on limiting the "fragmentation" that occurs when virtual machines (VM) are mapped to physical servers. Interestingly, they use a cooperative game theory approach to develop a cloud system with a sense of fairness in resource allocation. The system is supposed to be capable of providing on-demand resources, and creates a fairness-utilization trade off function to increase storage efficiency. In practice, this function allows servers to increase their efficiency. [15] distinguishes itself from past works, in terms of increased runtime efficiency, by spreading data between distinct processors and nodes rather than between the distributed clusters themselves. The author does so by presenting data partitioning algorithms prepared for matrix and linear algebra operations. Then, the partitionings are used in a "hierarchical manner" allowing "flexibility to be employed in a great range of problem domains and computational platforms". The resulting algorithm is an efficient means of partitioning large volumes of data. Significantly, it operates irrespective of the "power ratio that exists between the entities" and thereby minimizes execution time.

 Much like the preceding author, [14] discusses high-performance computing clusters, however develop an approach for integrating "data-intensive software framework with HPC cluster". In order to do so, they create a two-level storage system by combing Tachyon[22] with OrangeFS[23] to, respectively, combine an in-memory file system with a parallel file system. The benefit of which is an increase in the total throughput capacity of the system and high storage capacity. In summary, the above authors have shown several different methods of increasing the efficiency of data computation in the cloud, next I will discuss several applications of recent contributions to areas including language processing, compiling, and increasing smartphone efficiency.\\

%In contrast, authors in[5] attempt to increase cloud computation capacity by providing a Generic Framework for the allocation of data processing resources. In addition, they %claim to reduce the execution and migration time as well as network latency. %and is compared to Nephele.

\section{Relevant Applications}

 As mentioned above, the following discussion includes several recent applications of computation techniques. In the first article[7] the authors provide a parallelization techniques for membership tests specifically for Deterministic Finite Automata. They then apply these methods to identifying string patterns. To do so, they partition the data, match the subsets in parallel (using speculation), then combine the matching results. The DFA matches are "load-balanced on inhomogeneous parallel architectures such as cloud computing environments". The algorithm is particularly well suited, as it uses the structures of DFA to maintain semantics and avoid speed-downs. A shortened list of the applications of this research includes DNA sequence analyses, language scripting, and text editing.

 The work in [8] is also highly reliant on cloud infrastructure. To begin, the author proposes a cloud computation application to increase the efficiency of cell phones. As an example of cloud resource integration into common objects, apart from big data applications, difficulties in allocating resources both for smartphone users and for cloud systems is addressed. Using game theory and Nash equilibrium a decentralized computation-offloading scenario is formulated. In addition, it is pointed out that if many devices attempt to offload computation this would have the opposite effect of decreasing efficiency and gives the solution of a decentralized form of computation "such that mobile device users can make decisions locally". The product of which is a  program uses an offloading computation mechanism that transports data from mobile phones to cloud systems, thereby maximizing energy conservation and increasing the efficiency of cloud resource management.

 In contrast, authors in [9] present a compiler that retargets sequential Java code so that it may be run through Hadoop and implemented through the MapReduce paradigm, for parallel processing. Specifically, they rely on verified lifting to decrease the labor intensive and error prone task of programmers revisiting past programs. Bearing some resemblance to [7] authors in [13] process the data to find common structures. However, their application has notable distinguishing characteristics. Rather than assessing string, they locate features in videos. Specifically, they look for similarity in thousands of videos using the Pooled Time PoT Series approach. To overcome time and space constraints they create a Hadoop version of  PoT which involves splitting the data, and define Mapper and Reducer functions. The resulting system was then used to assist law enforcement in locating victims of human trafficking who had been sold on the dark web.

 To summarize, the above works provide  methods of allocating, partitioning, and combining data to increase efficiency of key user operations. The rest of the survey will focus on applications of data processing methods in health care, especially in regards to personalized medicine as it relates both to health monitoring and disease diagnosis.\\

\section{Applications in Health care} \label{SectionIV}
 The application of Big Data to human health problems has received significant attention lately. Used as a tool to reduce patient hospital visits, provide personalized medicine, and increase our understanding of the human body. The processing of this data, tends to be resource intensive. With continued development in this field, certain costs have decreased [28] and made personalized medicine accessible to some if not all [25]. The authors discussed below have attempted to increase the efficiency and efficacy of analyses while thereby reducing costs.

 Authors in [11] develop a model-based recursive partitioning procedure in a contribution to individualized health care. Specifically they use predictive factors to formulate these groups, and develop a segmented model of personalized treatment parameters for each group, and link these groups with a decision tree. In contrast, authors in [12] develop a recursive partitioning framework on competing risk data for predictive and prognostic models in cancer clinical trials. The authors in [10] present an algorithm, which combines conventional extraction with a feature importance filter. This relates to maintaining an expected amount of selected but irrelevant features during tasks of classification or regression, and addresses difficulties specific to machine-learning and data mining.\\

\subsection{Graph Theory}
 Graph theory, which has been used to develop networks in fields such as sociology and biology, is discussed for its application to neural networking in [17]. The authors contribute to the field of Integrated Information Theory of Consciousness by increasing the efficiency of calculating integrated information for neural data. They offer a Maximum Modularity Partition as opposed to comparable studies that use a Minimum Information Partition (MIP). Their method is "based on the graph-theoretic notion of modularity" and is an improvement over the MIP method in terms of cost efficiency and real data application.

Meanwhile, authors in [18] focus on the management of disease through mobile health platforms by offering an improvement in "optimal dynamic treatment regimes" of precision medicine patients. They accommodate an infinite time scale acting as a Markov decision process, with "minimal assumptions about the data-generating model". Further, to increase the efficacy of mobile health (mHealth) treatment management, they create a new reinforcement learning method to accommodate data obtained from mobile technologies. Thus, they offer "V-learning" as an alternative to the conventional Q-learning, which requires model free reinforcement [27], by instead selecting among a pre-specified class of policies.\\

\section{Applications in Health care Continued}
 Authors in [19] create a method of cancer survival prediction, from data obtained from RNA deep-sequencing. They cite one of the main challenges to analyzing this data as the "paucity of data samples compared to the high dimension of the expression profiles" which requires analysis that will effectively circumvent the considerable noise present in the data. In order to do so the authors propose a pipeline, supported by manifold learning and Laplacian support vector machines for unlabeled samples, the unlabeled samples are from data that did not follow up with patients. They take into account the variability in modality success and use a combinatory, "stacked generalization strategy" to increase the accuracy of the program.

 In contrast, authors in \cite{20,28} use multiple high-dimensional data matrices to create an algorithm to detect incidences of ovarian and liver cancer. They develop the algorithm by approximating the data matrices "with rank one outer products composed of sparse left singular vectors that are unique to each matrix and a right singular-vector" shared by each of the data matrices. The algorithm is then shown to operate more effectively than conventional standard singular value decomposition.\\

{\subsection{Data Analysis}}
 This section briefly discusses the issue of data useability, specifically data that, for some reason or another, requires significant processing before it can analyzed. In several of the above works[13],[19],[20], and [28]  authors have had to contend with noise, incompletion, and corruption in the data. There are many works that discuss established methods of filtering out noise and modeling incomplete data[30] which may or may not be detectable. For the purpose of this survey, undetectable problems in raw data will not be discussed.

 Authors in [16] devise two  algorithms that provide the informative columns of severely corrupted data. In addition, they demonstrate the functionality of this technique in the presence of sparse corru ption and outliers to the data. For a real world application, authors in \cite{24,29} address a common economic situation in which two entities have incomplete data regarding one another. Much like [4] and [8], they use game theory to make use of incomplete data. In this case however, they are analyzing competition between two vendors. They provide a simple learning algorithm for inventory competition which allows firms to make decisions without information about their competition due to the "implicit information encoded in their own total demand realizations affected  by their competitors' inventory decisions" that, using pure Markov strategies, results in equilibrium. Concluding remarks are provided below.\\

\section{CONCLUSIONS} \label{SectionVII}

 As can be seen, advancement in data analytic methods has had correspondingly wide effects in many fields of research and covers many more topics than those herein discussed. Topics that were lightly touched upon includes several papers related to cloud computing, implementations in health care: neural networks and cancer detection, economics, and work concerning user interfaces: smartphones, video, and text searches. In several of the above works, topics have dealt with imperfect and incomplete data. Solutions to which have employed game theory, Markov chains, and decision trees, to name a few. Though it may be quite obvious to the casual observer, it is hoped that by presenting the above works, further research concerning data analysis and its applications will be done.

\end{document}